\documentclass{emulateapj}
\usepackage{graphicx}
\usepackage[none]{hyphenat}
\usepackage{amsmath}
\usepackage{booktabs}
\usepackage{multirow}
\usepackage{txfonts}
\usepackage{enumitem}
\usepackage{amsmath}
\usepackage{algorithmic}
\usepackage{bm}
\usepackage{mathrsfs}
\usepackage{color}
\usepackage{lineno}

\usepackage[colorlinks,linkcolor=blue,anchorcolor=blue,citecolor=blue]{hyperref}

\begin{document}

\title{Studying the properties of compressible MHD turbulence by synchrotron fluctuation statistics}
\author{Ru-Yue Wang\altaffilmark{1}, Jian-Fu Zhang\altaffilmark{1}, Alex Lazarian\altaffilmark{2,3}, Hua-Ping Xiao\altaffilmark{1}, Fu-Yuan Xiang\altaffilmark{1}}
\email{jfzhang@xtu.edu.cn (JFZ), hpxiao@xtu.edu.cn (HPX)}
\altaffiltext{1}{Department of Physics, Xiangtan University, Xiangtan, Hunan 411105, China}
\altaffiltext{2}{Department of Astronomy, University of Wisconsin, 475 North Charter Street, Madison, WI 53706, USA}
\altaffiltext{3}{Centro de Investigación en Astronomía, Universidad Bernardo O’Higgins, Santiago, General Gana 1760, 8370993, Chile}

\begin{abstract}

We study the observable properties of compressible MHD turbulence covering different turbulence regimes, based on synthetic synchrotron observations arising from 3D MHD numerical simulations. Using the synchrotron emissivity and intensity, we first explore how the cosmic ray spectral indices affect the measurements of turbulence properties by employing normalized correlation functions. We then study how the anisotropy of synchrotron total and polarization intensities arising from three fundamental MHD modes vary with the viewing angle, i.e., the angle between the mean magnetic field and the line of sight. We employ the ratio of quadrupole moment to the monopole one (QM) for this purpose. Our numerical results demonstrate that: (1) the two-point correlation function of synchrotron statistics for the arbitrary cosmic ray spectral index is related to the special case of magnetic field index $\gamma=2$ in agreement with the analytical formulae provided by Lazarian \& Pogosyan (2012); (2) the anisotropy of synchrotron total and polarization intensities arising from Alfv\'en and slow modes increases with the increase of the viewing angle, while that of fast mode remains almost unchanged with the viewing angle; (3) the analytical formulae of synchrotron intensities for studying turbulence can be applied to describing statistics of polarization intensities, and the QM can be successfully used to recover turbulence anisotropy. This study validates Lazarian \& Pogosyan's analytical approach and opens a way to study turbulence from observations.
\end{abstract}


\section{Introduction}
Magnetohydrodynamic (MHD) turbulence, resulting from the interaction of fluids and magnetic fields, is ubiquitous in astrophysical environments. 
The most significant evidence is the spectral distribution of electron density fluctuations in the Milky Way, termed as big power law in the sky (\citealt{Armstrong1995ApJ...443..209A, Chepurnov2010ApJ...710..853C}). MHD turbulence has significant impact on fundamental astrophysical processes, such as star formation (\citealt{MacLow2004RvMP...76..125M, McKee2007ARA&A..45..565M, Crutcher2012ARA&A..50...29C}), propagation and acceleration of cosmic rays (\citealt{Yan2008ApJ...673..942Y, Xu2018ApJ...868...36X}), heat conduction (\citealt{Narayan2001ApJ...562L.129N, Lazarian2006ApJ...645L..25L}), turbulent magnetic reconnection (\citealt{Lazarian1999ApJ...517..700L, Lazarian2020PhPl...27a2305L}). Therefore, a comprehensive understanding of the properties of MHD turbulence is indispensable to describing astrophysical processes.

The theory of MHD turbulence has been developing for several decades (see \citealt{Biskamp2003matu.book.....B} for a book; \citealt{Beresnyak2019tuma.book.....B} for a recent book). A key turning point in the construction of modern MHD turbulence theory can date back to the study of \cite{Goldreich1995ApJ...438..763G}, which focuses on incompressible MHD turbulence. The most important contribution of GS95 theory is the prediction of the scale-dependent anisotropy of turbulence cascade. Further insight into the nature of MHD turbulence cascade was provided by the theory of turbulent reconnection 
in \cite{Lazarian1999ApJ...517..700L}. Here, it was shown that the MHD turbulence can be presented as a way of eddies for which rotation is aligned with the direction of the magnetic field in their vicinity. Indeed, the prediction of LV99 is that the turbulent reconnection takes place within one eddy turnover time and therefore the motion of magnetic field lines perpendicular to the local magnetic field is not constrained by magnetic field tension. Thus the turbulent energy cascades along the path of the least resistance, which involves mixing magnetic field lines rather than bending them. 
It is worth noting that from the picture of MHD turbulence, the anisotropic relation was originally proposed by GS95 in the frame of the mean magnetic field, which should in fact be relevant only in the reference frame related to the magnetic field associated with the turbulent eddies, i.e., the local frame of reference (LV99, \citealt{Cho2000ApJ...539..273C, Maron2001ApJ...554.1175M}).

The decomposition of compressible MHD turbulence into three modes is an important topic in the development process of MHD turbulence theory (\citealt{Cho2002PhRvL..88x5001C, ChoLazarian2003MNRAS.345..325C, Kowal2010ApJ...720..742K, Wang2021MNRAS.505.6206W, Hernandez-Padilla2020ApJ...901...11H, Hu2021ApJ...915...67H, Zhang2021ApJ...922..209Z}). The research results have become the important components of modern MHD turbulence theory. Especially, MHD turbulence is numerically decomposed into Alfv{\'e}n, slow and fast modes in the Fourier space (CL02; CL03), which provides a new perspective for understanding MHD turbulence. Later, the Fourier decomposition was confirmed by the wavelet decomposition, and generalized to the cases of no mean magnetic field and super-Alfv{\'e}nic turbulence (\citealt{Kowal2010ApJ...720..742K}).

Although direct numerical simulation has achieved many valuable results on the properties of MHD turbulence, the current simulation up to Reynolds number $R_{\rm e}\simeq 10^5$ still cannot simulate realistic astrophysical environments (\citealt{Beresnyak2019LRCA....5....2B}) given the inherent high Reynolds number characteristics of astrophysics, such as ISM with $R_{\rm e}>10^{10}$. Therefore, a new observation-based research perspective has been adopted to attempt to avoid the difficulties of direct numerical simulation. 

Currently, a number of statistical techniques from an observational perspective have been developed to explore the properties of MHD turbulence. Those can be divided into two major categories in terms of information obtained. One involves the Doppler-shifted spectroscopic data (see \citealt{Lazarian2009SSRv..143..357L} for a review; \citealt{Lazarian2000ApJ...537..720L, Chepurnov2009ApJ...693.1074C, Kandel2016MNRAS.461.1227K, Kandel2017MNRAS.464.3617K}). The other important branch of the research is related to synchrotron radiation (e.g., \citealt{Lazarian2012ApJ...747....5L, Lazarian2016ApJ...818....178L, Herron2018ApJ...853....9H, Zhang2022FrASS...9.9370Z}), the most typical of which are LP12 and LP16.

When the relativistic electrons spiraling about the magnetic field are accelerated, a synchrotron radiation signal is emitted. Analyzing synchrotron radiation fluctuations provides a powerful way of studying the properties of astrophysical magnetic fields (\citealt{Rickett1990ARA&A..28..561R, Heiles2012SSRv..166..293H, Hill2015AAS...22512705H, Haverkorn2019Galax...7...26H, Thomson2019MNRAS.487.4751T, Wolleben2021AJ....162...35W, Erceg2022A&A...663A...7E}). LP12 first proposed a theoretical description of synchrotron intensity fluctuations arising from magnetic turbulence. They provided analytical formulae that relate the correlation of synchrotron fluctuations for an arbitrary index of relativistic electrons to the correlation for a particular magnetic field index $\gamma=2$ and predicted the anisotropy of synchrotron intensity by the ratio between quadrupole and monopole parts, being sensitive to the compressibility of underlying turbulence. Furthermore, based on a statistical analysis of synchrotron polarization intensity, LP16 suggested polarization spatial and polarization frequency analysis techniques could reveal the properties of MHD turbulence, such as spectrum, anisotropy, and correlation scales.

The polarization frequency analysis technique was studied by \cite{Zhang2016ApJ...825..154Z}, who confirmed the theoretical predictions of LP16 and achieved the measurement of the spectral index of underlying magnetic turbulence. Subsequently, the technique was applied to the polarization observation of optical/infrared blazars (\citealt{Guo2017ApJ...843...23G}). Based on synthetic and real simulation data of MHD turbulence, the polarization spatial analysis technique was successfully tested in the
spatially coincident synchrotron emission and Faraday regions (\citealt{Lee2016ApJ...831...77L}) and the spatially separated regions (\citealt{Zhang2018ApJ...863..197Z}). These efforts focused on studying how to recover the spectral index of magnetic turbulence.   

Using the quadrupole ratio method proposed in the synchronized intensity fluctuation study (LP12), \cite{Lee2019ApJ...877..108L} numerically studied the statistical description of anisotropy of polarized synchrotron intensity arising from one spatial region and two spatially separated regions, respectively. With the same method, \cite{Wang2020ApJ...890...70W} explored the compressibility and anisotropy of MHD turbulence and found that the anisotropic features of Alfv\'en, slow and fast modes are in agreement with the earlier direct numerical simulation of MHD turbulence (CL02; CL03).

In addition, synchrotron intensity
and polarization gradient techniques (\citealt{Lazarian2017ApJ...842...30L, Lazarian2018ApJ...865...59L}) have been developed for tracing the direction of the magnetic field (\citealt{Zhang2019MNRAS.486.4813Z, Zhang2019ApJ...886...63Z, Zhang2020ApJ...895...20Z, Wang2021MNRAS.505.6206W}). It is noticed that synchrotron polarization gradient was suggested by \cite{Gaensler2011Natur.478..214G} to constrain the sonic Mach number of warm, ionized ISM. Later, various polarization diagnostic quantities were derived in \cite{Herron2018ApJ...853....9H} and applied to simulation and observational data for distinguishing between backlit and internal emission (\citealt{Herron2018ApJ...855...29H}). Applying synchrotron diagnostic gradients to the archive data from the Canadian Galactic Plane Survey (\citealt{Taylor2003AJ....125.3145T, Zhang2019ApJ...886...63Z}) made consistent predictions for the gradient directions and the Galactic magnetic field directions. More recently, \cite{Wang2021MNRAS.505.6206W} explored the capabilities of gradient techniques in the case of spatial separation of synchrotron polarization and Faraday rotation regions.

The purpose of this paper is to advance the study of the compressibility of MHD turbulence. We want to know whether the analytical descriptions provided in LP12 can be used to reveal the anisotropy of magnetic turbulence. Does the change of the angle between the mean magnetic field{\footnote{The mean magnetic field is 
consistent with the large-scale or ordered magnetic field (\citealt{Fletcher2011MNRAS.412.2396F}). The presence of the magnetic field makes MHD turbulence anisotropic (\citealt{Montgomery1981anisotropic, Shebalin1983JPlPh..29..525S, Higdon1984ApJ...285..109H}; GS95), and the larger the mean magnetic field, the more pronounced the anisotropy. In the system of reference of the mean magnetic field, the anisotropy of eddies is scale-independent (different from the anisotropy of GS95) and the degree of anisotropy is determined by the largest eddies (\citealt{Cho2002ApJ...564..291C, Esquivel2005ApJ...631..320E}).}} and the line of sight hinder the application of synchrotron statistics techniques? Does the spectral index of relativistic electrons affect the measurement of compressible turbulence properties, compared with the analytical descriptions of LP12? At the same time, the studies of synchrotron radiation intensity are also generalized to the synchrotron polarization one.

This paper is organized as follows. In Section 2, we provide theoretical descriptions including the fundamental theory of MHD turbulence, the calculation of synchrotron radiation, the analytical expressions of LP12, and statistical methods. In Section 3, we introduce the procedure of numerical simulation of MHD turbulence. The results are presented in Section 4. Finally, we make a discussion and summary in Sections 5 and 6, respectively.

\begin{deluxetable*}{cccccccccccc}
\tabletypesize{\scriptsize}
\tablecaption{
Data cubes with the numerical resolution of $512^3$ arise from compressible MHD turbulence simulations. Different turbulence regimes are mainly characterized by the parameters: Alfv\'enic Mach number $M_{\rm A}$, sonic Mach number $M_{\rm s}$, and the plasma parameter $\beta=2M_{\rm A}^2/M_{\rm s}^2$. Other parameters are listed as follows----$B_0$ the initial magnetic field strength; $\delta B_{\rm rms}$: root mean square of the random magnetic field; ${\langle B \rangle}$: regular magnetic field; $L_{\rm inj}$: injection scale; $l_{\rm 
b}$: correlation length of magnetic field; $l_{\rm v}$: correlation length of velocity; $L_{\rm trans}$:  transition scale for $M_{\rm A}<1$; $L_{\rm A}$: transition scale for $M_{\rm A}>1$; and $l_{\rm diss}$: dissipation scale. }
\tablewidth{170mm}
\tablehead{\colhead{run} 
   & \colhead{Init $B_0$}
   & \colhead{$M_{\rm s}$} 
   & \colhead{$M_{\rm A}$}
   & \colhead{$\beta$}
   & \colhead{$\delta B_{\rm rms}/{\langle B \rangle}$}
   & \colhead{$L_{\rm inj}$}
   & \colhead{$l_{\rm b}$}
   & \colhead{$l_{\rm v}$}
   & \colhead{$L_{\rm trans}(L_{\rm A})$}
   & \colhead{$l_{\rm diss}$}
}
\startdata
1  & 1.0  & 9.92  & 0.50  & 0.005  & 0.465  & $\sim0.4L$  & $\sim0.33L$ & $\sim0.33L$ &$\sim51.20$ & $\sim0.025L$ \\
2  & 1.0  & 4.46  & 0.55  & 0.030  & 0.467  & $\sim0.4L$  & $\sim0.33L$ & $\sim0.25L$ &$\sim61.95$ & $\sim0.025L$  \\
3  & 1.0  & 3.16  & 0.58  & 0.067  & 0.506  & $\sim0.4L$  & $\sim0.33L$ & $\sim0.25L$ &$\sim68.89$ & $\sim0.025L$  \\
4  & 1.0  & 0.87  & 0.70  & 1.295  & 0.579  & $\sim0.4L$  & $\sim0.33L$ & $\sim0.33L$ &$\sim100.35$ & $\sim0.025L$  \\
5  & 1.0  & 0.48  & 0.65  & 3.668  & 0.614  & $\sim0.4L$  & $\sim0.33L$ & $\sim0.20L$ &$\sim86.52$ & $\sim0.025L$ \\
6  & 0.1  & 3.11  & 1.69  & 0.591  & 5.254  & $\sim0.4L$  & $\sim0.20L$ & $\sim0.33L$ &$\sim42.43$  & $\sim0.031L$  \\
7  & 0.1  & 0.45  & 1.72  & 29.219  & 6.345  & $\sim0.4L$  & $\sim0.20L$ & $\sim0.20L$ &$\sim40.25$  & $\sim0.031L$ \\
\enddata
\label{table_1}
\end{deluxetable*}

\section{Theoretical Descriptions}

\subsection{Fundamentals of MHD turbulence theory} \label{MHD_basic_theory}
GS95 theory is a starting point for understanding modern MHD turbulence. It focused on incompressible, trans-Alfv{\'e}nic turbulence with Mach number $M_{\rm A}=V_{\rm L}/V_{\rm A}=1$, where $V_{\rm A}$ and $V_{\rm L}$ are  Alfv{\'e}nic velocity and injection velocity $V_{\rm L}$ at the injection scale $L_{\rm inj}$, respectively. With an assumption of the critical balance, $v_{\rm k}k_{\perp}=V_{\rm A} k_{\parallel}$, GS95 predicted the anisotropic scaling relation with regard to the parallel and perpendicular wave numbers, namely $k_{\parallel}\varpropto k_{\perp}^{2/3}$, where $v_{\rm k}$ is the velocity at the scale $k^{-1}$.

Later, GS95 theory was generalized to the cases for both $M_{\rm A}<1$ (LV99) and $M_{\rm A}>1$ (\citealt{Lazarian2006ApJ...645L..25L}), respectively. For the former, the injection velocity $V_{\rm L}$ is smaller than Alfv{\'e}nic velocity $V_{\rm A}$, called sub-Alfv{\'e}nic turbulence. 
When driving at an injection scale $L_{\rm inj}$, the turbulence is weak from the injection scale $L_{\rm inj}$ to the transition scale 
\begin{equation}
L_{\rm trans}=L_{\rm inj}M_{\rm A}^{2} .
\end{equation}
The turbulence becomes strong when the scale is smaller than the transition scale $L_{\rm trans}$ but larger than the dissipation scale $l_{\rm diss}$. In this case, the relationship between parallel and perpendicular scales for an eddy is described as:  
\begin{equation}
l_{\|}\approx L_{\rm inj}^{1/3}l_{\perp}^{2/3}M_{\rm A}^{-4/3}. \label{anis}
\end{equation}
In terms of perpendicular motion, the spectrum of Alfv{\'e}n modes for $l_{\bot}<L_{\rm trans}$ is Kolmogorov, which is self-evident from the eddy description of Alfv{\'e}nic turbulence provided above. Compared to the trans-Alfv{\'e}nic case{\footnote{Formally, the elongation of the eddies corresponds to trans-Alfv\'enic turbulence but injected at the scale $L_{\rm eff}=L_{\rm inj} M_A^{-4}$ (\citealt{Lazarian2021ApJ...923...53L}).}, the eddies for sub-Alfv{\'e}nic turbulence are more elongated.}
When the spectrum is measured in the global system, i.e., the system of reference regarding the mean magnetic field, the effect of perpendicular motions is dominant by the case both parallel to the mean magnetic field and perpendicular to it. 

If $M_{\rm A}>1$, the turbulence is super-Alfv{\'e}nic.
Such turbulence at the scale close to $L_{\rm inj}$ has an essentially hydrodynamic Kolmogorov property. The cascade becomes fully magnetic from hydrodynamic turbulence to MHD one at the scale (\citealt{Lazarian2006ApJ...645L..25L}):
\begin{equation}
L_{\rm A}=L_{\rm inj}M_{\rm A}^{-3} ,
\end{equation}
which corresponds to the scale at which turbulent velocity gets equal to the Alfv\'en velocity. In the range of $[L_{\rm A},l_{\rm diss}]$, the super-Alfv{\'e}nic turbulence is similar to trans-Alfv\'enic MHD turbulence with the injection scale equal to $L_{\rm A}$.

GS95 theory provided insight into expected properties of compressible MHD turbulence (see also \citealt{Lithwick2001ApJ...562..279L}).
The incompressible Alfv{\'e}n and compressible slow modes have the same scale-dependent anisotropy as the Alfv{\'e}n modes in incompressible turbulence (see above). The Alfv{\'e}n modes also impose the Kolmogorov property inducing turbulence complexity to a simple formula $E(k) \propto k^{-5/3}$, where $E$ is the power spectrum of turbulent motions, while the scaling of compressible fast mode is less clear. It is suggestive that for subsonic driving the spectrum follows $E(k) \propto k^{-3/2}$ (CL03), while for supersonic driving the spectrum steepens to $k^{-2}$. In addition, the Alfv{\'e}n and slow modes show GS95 scale-dependent anisotropy, while the fast mode with isotropy is different from the above two modes.

\begin{figure*}
\centering
\includegraphics[width=1.0\textwidth,height=0.6\textheight]{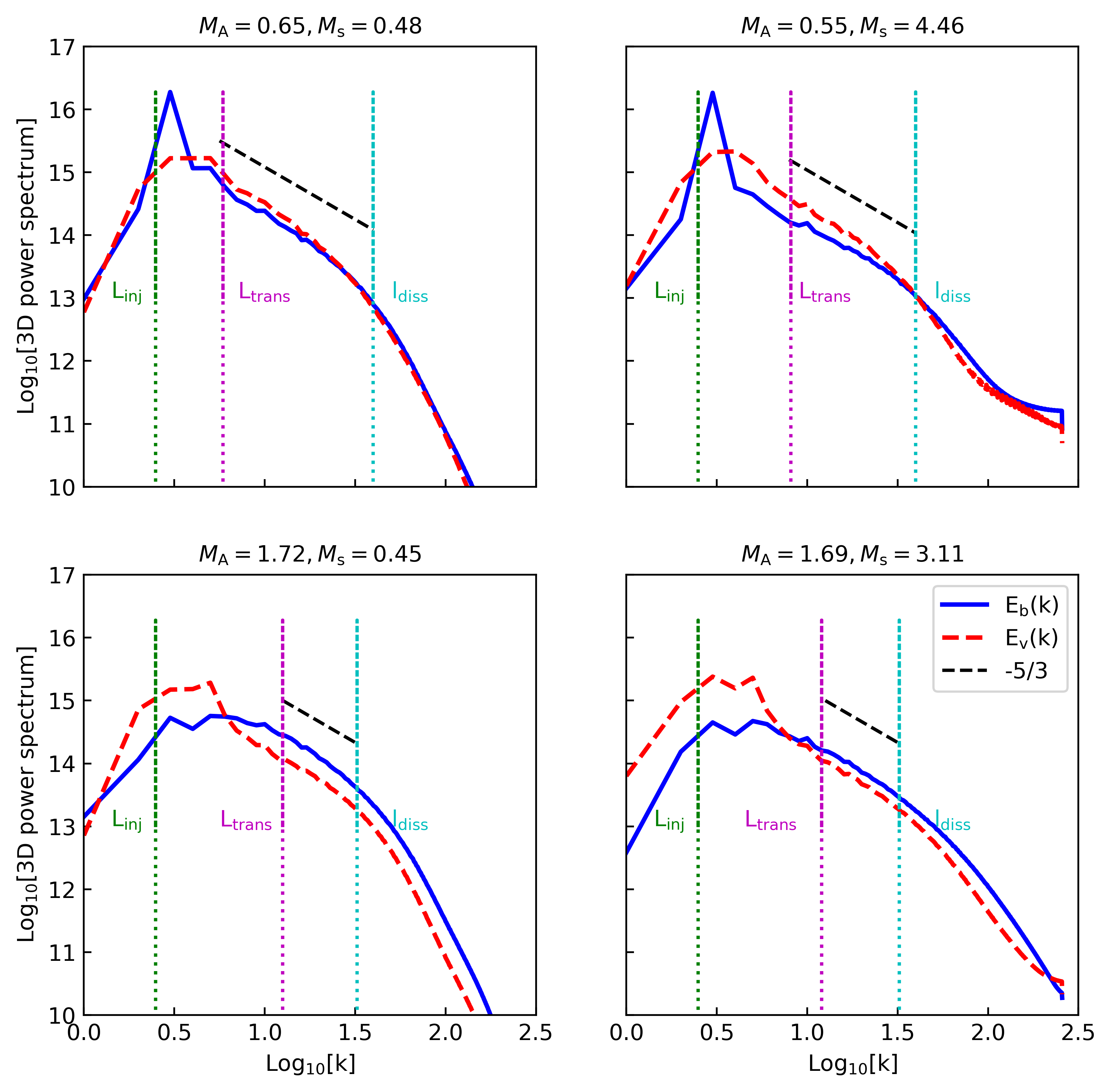}
\caption{{The power spectra of the magnetic field in different turbulence regimes. The green, magenta, and cyan-dotted vertical lines denote the injection, transition, and dissipation scales of turbulence, respectively.}}
\label{fig:power} 
\end{figure*}

\begin{figure*}
\centering
\includegraphics[width=1.0\textwidth,height=0.6\textheight]{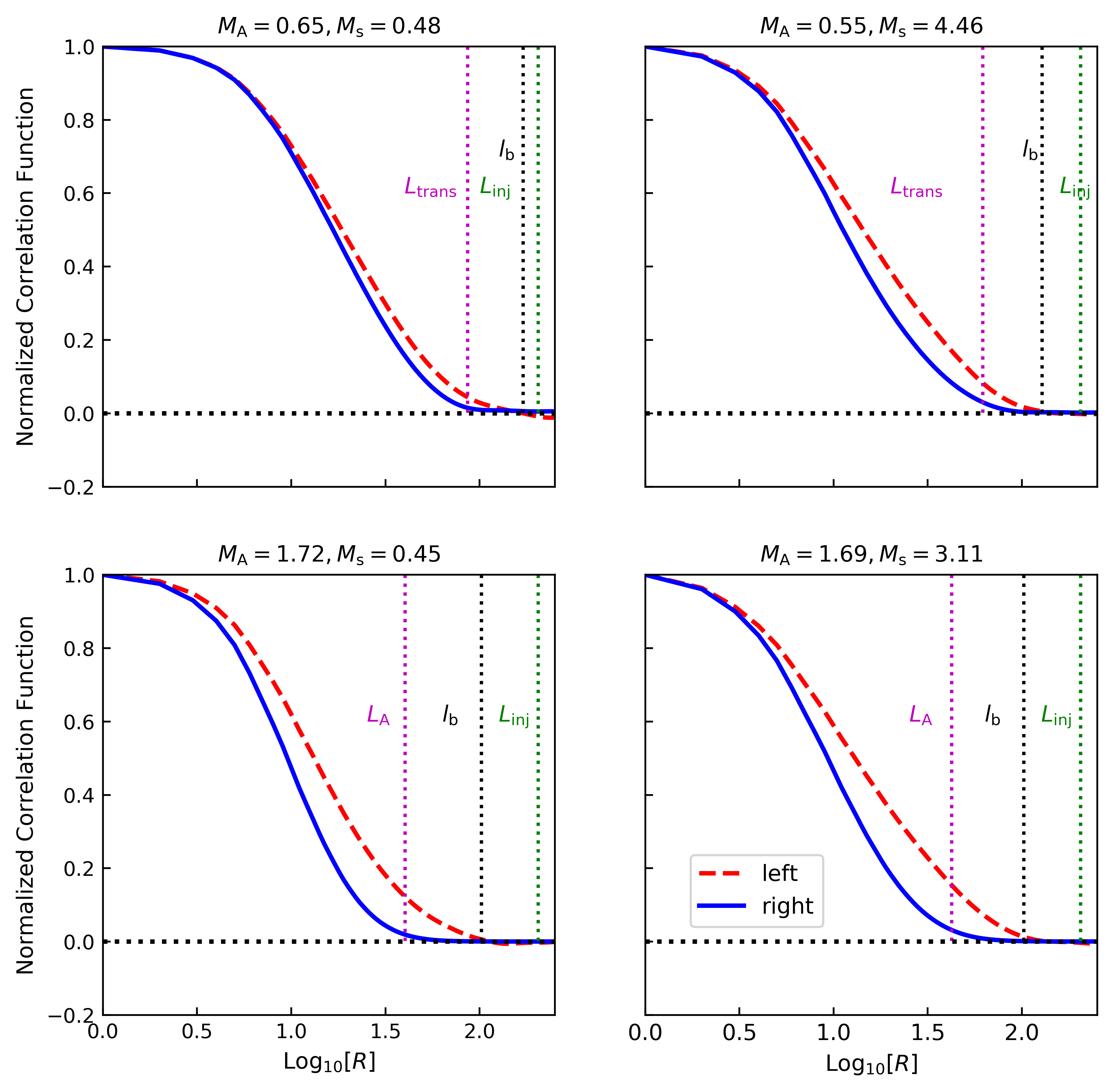}
\caption{{The numerical test of correlation's analytical formula in different turbulence regimes. The red dashed lines represent the results calculated by the Equation (\ref{ncf}) for the magnetic field index $\gamma=$2, and the blue solid lines on the right side of Equation (\ref{ccc}), i.e., the square of Equation (\ref{B_y}). The green and magenta dotted vertical lines denote the injection and transition scales of turbulence, respectively. The black dotted vertical line denotes the correlation length of the magnetic field.}}
\label{fig:left_right} 
\end{figure*}

\begin{figure*}
\centering
\includegraphics[width=1.0\textwidth,height=0.6\textheight]{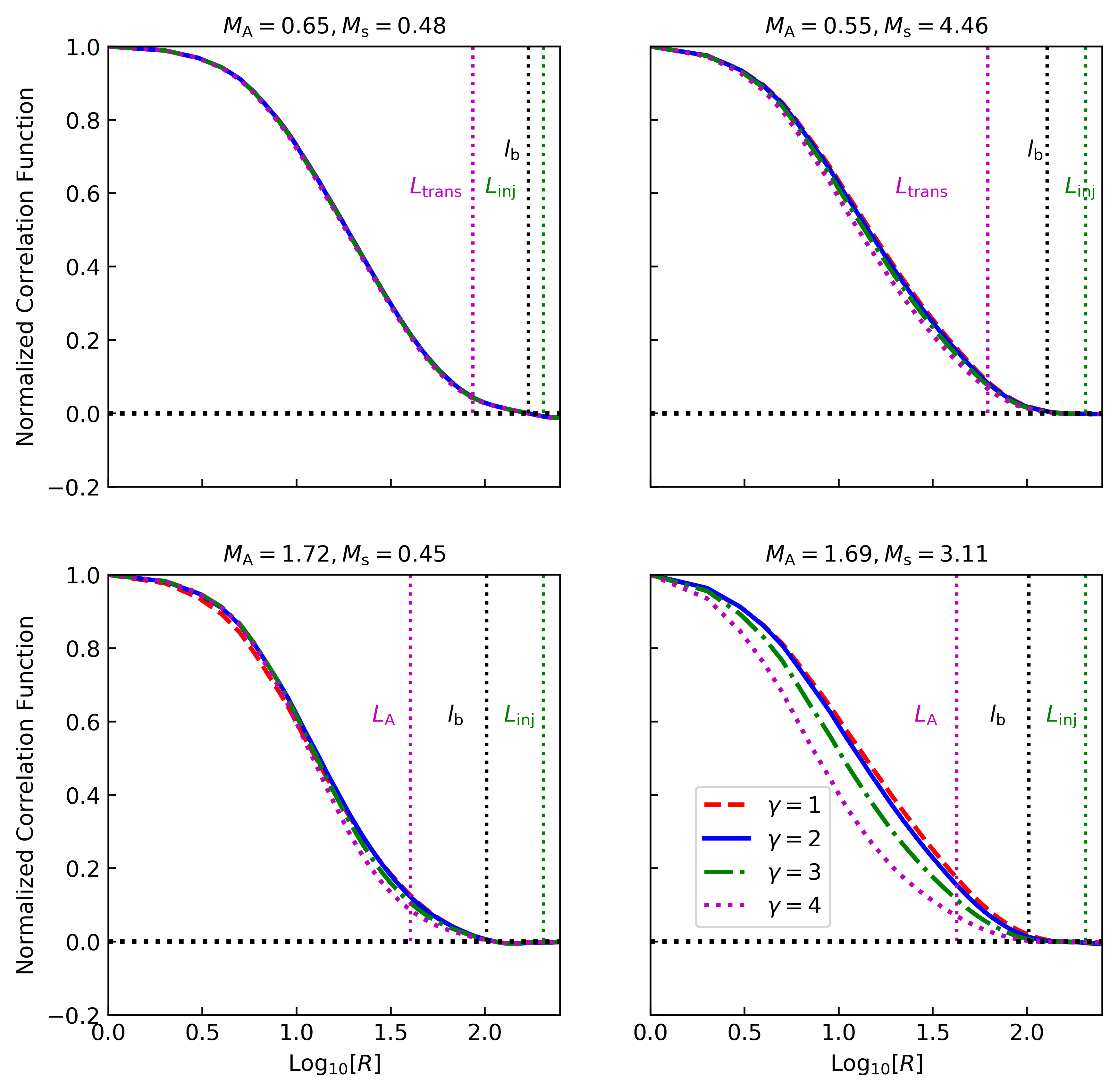}
\caption{The NCFs of synchrotron emissivity for varying electron indices in different turbulence regimes. The green and magenta dotted vertical lines denote the injection and transition scales of turbulence, respectively. The black dotted vertical line denotes the correlation length of the magnetic field.
}\label{fig:ncf_total} 
\end{figure*}

\subsection{Synchrotron Radiative Processes in the Magnetic Turbulence}\label{SecRad}
The magnetic field and relativistic electron energy distribution are two major factors that determine synchrotron radiation. 
In this paper, we assume that the relativistic electron population has an isotropic pitch angle distribution and a power-law energy distribution\footnote{We do not simulate the propagation of cosmic ray electrons. 
Therefore, we do not address the issue of the hypothetical correlation of their density with 
magnetic field strength. In this case, the cosmic ray electron is uncorrelated with the magnetic field  (\citealt{BeckShukurov2003A&A...411...99B, Beck2005AN....326..414B, Seta2019Galax...7...45S}).}, described by
\begin{equation}
N(E) dE=N_{0}E^{2\alpha-1}dE, \label{eq:2}
\end{equation}
where $N(E)$ is the number density of relativistic electron with energy between $E$ and $E+dE$, and $N_{0}$ a normalization constant and $\alpha$ the spectral index.

The synchrotron radiation intensity at a fixed frequency $\nu$ is calculated by (\citealt{Ginzburg1965ARA&A...3..297G, Ginzburg1981tpas.book.....G})
\begin{eqnarray}
I({\bm X})=&&  \frac{e^3}{4\pi m_{\rm e} c^2} \int_0^L \frac{\sqrt 3}{2-2\alpha}
\Gamma \left(\frac{2-6\alpha}{12}\right)\Gamma \left(\frac{22-6\alpha}{12}\right)
\nonumber \\
&&  \times \left(\frac{3e}{2\pi m_{\rm e}^3 c^5}\right)^{-\alpha} 
N_{0}B_{\perp}^{1-\alpha}({\bm X}, z) \nu^{\alpha}dz, 
\label{eq:I}
\end{eqnarray}
where ${\bm X}=(x,y)$ represents a 2D vector in the plane of the sky, $\Gamma$ the gamma function, $\gamma=1-\alpha$ the index of the magnetic field, $B_{\perp}$ the magnetic field component perpendicular to the line of sight (LOS), and $L$ the emitting-region size. The other symbols have their usual meanings.

When the radiation is linearly polarized, the intrinsic polarization intensity is calculated by 
\begin{equation}
P_{0}(\bm X)=\Pi_{L}I(\bm X), \label{eq:Pin}
\end{equation}
where $\Pi_{L}=\frac{3-3\alpha}{5-3\alpha}$ is fraction polarization degree. The observable Stokes parameters $Q$ and $U$ related to the polarization angle $\Psi$ can be expressed as $Q(\bm X)=P_{0}(\bm X)\cos2\Psi$ and $U(\bm X)=P_{0}(\bm X)\sin2\Psi$, respectively. 
When the polarized emission does not experience the Faraday rotation effect, the polarization angle corresponds to the intrinsic polarization angle, namely $\Psi=\Psi_{0}={\pi/2}+\arctan({B_{\rm y}}/{B_{\rm x}})$. 
When encountering with Faraday rotation effect, the polarization angle is written as $\Psi=\Psi_{0}+{\rm RM} \lambda^2$, with the Faraday rotation measure 
\begin{equation}
{\rm RM} ({\bm X}, z) =0.81\int_{0}^{z}{n_{\rm e}({\bm X}, z^{'})} B_{\parallel}({\bm X}, z^{'}) dz^{'}~ \rm rad~m^{-2}, \label{eq:RM}
\end{equation}
where $n_{\rm e}$ is the number density of thermal electron, $B_{\parallel}$ the magnetic field component along the LOS, and $z$ a variable. The integration of Equation \ref{eq:RM} is along the LOS from the observer to the position of the source at $z^{'}=z$. Moreover, the observable polarization intensity can be expressed by
\begin{equation}
 PI(\bm X)=\sqrt{Q^{2}(\bm X)+U^{2}(\bm X)}. \label{eq:PI}
\end{equation}

\subsection{Analytical expressions of correlation for anisotropic turbulence}\label{ACorre}
LP12 provided the expressions of the correlation function of synchrotron intensity, which was related to the change in the relativistic electron spectral index for isotropic and anisotropic turbulence, respectively. The formula for isotropic turbulence has been tested by numerical simulation (\citealt{Herron2016ApJ...822...13H}). 
The anisotropic turbulence will be explored in the paper. 
In this case, the normalized correlation function (NCF) of synchrotron emissivity is defined by
\begin{equation}
\widetilde{\rm CF}=\frac{\langle B_{\perp}^{\gamma}{(\bf x)}B_{\perp}^{\gamma}({\bf x+r})\rangle-{\langle B_{\perp}^{\gamma}({\bf x)}\rangle^2}}{\langle B_{\perp}^{\gamma}{(\bf x)}^2\rangle-\langle B_{\perp}^{\gamma}{(\bf x)}\rangle^{2}}, \label{ncf}
\end{equation}
where $\langle...\rangle$ indicates an average through the whole volume space; $B_{\perp}(\bf x)=\sqrt{\langle{B_{\rm x}}\rangle^2+{B_{\rm y}}^2({\bf x})}$ is the component of the magnetic field perpendicular to the LOS; ${\bf x} = (x,y,z)$ represents the position vector of any spatial point in the emitting region; $\bf r$ is the separation vector between two points; and $\langle B_{\rm x}\rangle$ indicates the mean magnetic field.

The index $\gamma$ was a big problem for constructing the statistical theory of synchrotron fluctuations. This problem was resolved by LP12, which obtained analytical relation between the correlations of synchrotron fluctuations for an arbitrary $\gamma$ and those for $\gamma=2$.
The relations obtained in LP12 demonstrated that the two types of statistics differ by a factor that is independent of the lag over which the correlations were measured. This opened a way to study the synchrotron statistics for $\gamma=2$, simplifying the problem significantly.

Setting $\gamma=2$, Equation (\ref{ncf}) can be expressed as (see LP12)
\begin{equation}
\widetilde{\rm CF}\simeq c ({\bf r})^2, \label{ccc}
\end{equation}
where the symbol $c(\bf r)$ on the right side is 
\begin{equation}
c({\bf{r}})=\frac{\langle{B_{y}({\bf x})}{B_{y}({\bf x+r})}\rangle-\langle{B_{y}({\bf x})}\rangle^2}{\langle{B_{y}({\bf x})^2}\rangle-\langle{B_{y}({\bf x})}\rangle^2}. \label{B_y}
\end{equation}
Since the magnetic field is assumed to be far from fluctuations along the mean magnetic field direction, i.e., $x$-axis direction, Equation (\ref{ccc}) is only associated with the component $B_{\rm y}$.

\subsection{Analytical expressions of anisotropy for three modes}\label{Aaniso}
Here, we briefly summarize analytical expressions for the anisotropy of three modes as follows (see LP12 for more details).

{\it Alfv{\'e}n mode}: The multipole of synchrotron (polarization) intensity is expressed as  
\begin{eqnarray}
&&\tilde{\mathcal D}_{\rm m}(R)\approx {C}_{\rm m}(2/3)~[{W}_{\rm I}~(\hat{\mathcal P}_{\rm m}-\frac{1}{2}\epsilon~(\hat {\mathcal P}_{\rm m+2}+\hat{\mathcal P}_{\rm m-2}))
\nonumber\\
&&+{W}_{\rm L}\sum_{n=-\infty}^{\infty}[\hat{\mathcal P}_{\rm n}-\frac{1}{2}\epsilon~(\hat {\mathcal P}_{\rm n+2}+\hat{\mathcal P}_{\rm n-2})]~{A}_{\rm m-n}^{A}(\theta)]~R^{5/3}, \label{amono}
\end{eqnarray}
where the multipole includes two parts associated with the weight function $W_{\rm I}$ and $W_{\rm L}$.
The relevant parameters in Equation (\ref{amono}) are introduced as follows:

\noindent{(1)} The parameter $C_{\rm m}(2/3)$, depending on the scaling of synchrotron correlation ($\mu$), can be determined by
\begin{equation}
C_{\rm m}(\mu)=-\frac{i^{m}~\Gamma[\frac{1}{2}(|m|-\mu-1)]}{2^{2+\mu}~\Gamma[\frac{1}{2}(|m|+\mu+3)]}.
\end{equation}

\noindent{(2)} The isotropic and local weight functions, written as (\citealt{Lazarian2022ApJ...935...77L})
\begin{equation}
W_{\rm I}\approx\frac{M_{\rm A}^{2}}{2+2M_{A}^2}, \label{wi}
\end{equation} 
\begin{equation}
W_{\rm I}+W_{\rm L}\approx\frac{1+M_{\rm A}^2}{1+2M_{\rm A}^2}, \label{wl}
\end{equation}
characterize the relative proportion of the degree of isotropy and anisotropy, respectively.

\noindent{(3)} The parameter $\epsilon$, representing the level of the anisotropy, is defined by 
\begin{equation}
\epsilon=\frac{\sigma_{\rm xx}-\sigma_{\rm yy}}{\sigma_{\rm xx}+\sigma_{\rm yy}},\label{aniscoeff}
\end{equation}
where $\sigma_{\rm xx}=\sigma_{\parallel}^{2}\sin^{2}\theta+\frac{1}{2}\sigma_{\perp}^{2}\cos^{2}
\theta$, 
$\sigma_{\rm yy}=\frac{1}{2}\sigma_{\perp}^{2}$. 
Here, $\theta$ is the angle between the mean magnetic field and LOS, $\sigma_{\parallel}$ and $\sigma_{\perp}$ denote the fluctuations parallel and perpendicular to the mean magnetic field, respectively.

\noindent{(4)} The harmonic decomposition of 2D spectra is defined by
\begin{eqnarray}
\hat{{\mathcal P}_{\rm m}}=\frac{1}{2\pi} \int_0^{2\pi}d\psi e^{-im\psi}{{ \exp}[-{M}_{\rm A}^{-\frac{4}{3}}\frac{{\left|\cos\psi\right|} \sin\theta}{(1-\cos^2\psi\sin^2\theta)^{\frac{2}{3}}}]},
\end{eqnarray}
and $\hat{{\mathcal P}_{\rm n}}$ has the same expression as $\hat{{\mathcal P}_{\rm m}}$ with uncertain subscripts $n$.

\noindent{(5)} The geometrical function for Alfv{\'e}n mode is given by
\begin{equation}
{A}_{\rm m-n}^{A}(\theta)=\frac{1}{2\pi}\int_0^{2\pi}d\psi e^{-i(m-n)\psi}\frac{\cos^2\theta}{1-\cos^2\psi\sin^2\theta}.
\end{equation}

{\it Slow mode}: The multipole of synchrotron fluctuations is defined by
\begin{eqnarray}
&&\tilde{\mathcal D}_{\rm m}(R)\approx C_{\rm m}(2/3)~[W_{\rm I}~(\hat{\mathcal P}_{\rm m}-\frac{1}{2}\epsilon~(\hat {\mathcal P}_{\rm m+2}+\hat{\mathcal P}_{\rm m-2}))
\nonumber\\
&&+W_{\rm L}{\sin^2\theta}\sum_{n=-\infty}^{\infty}[\hat{\mathcal P}_{\rm n}-\frac{1}{2}\epsilon~(\hat {\mathcal P}_{\rm n+2}+\hat{\mathcal P}_{\rm n-2})]~{A}_{\rm m-n}^{S}(\theta)]~R^{\frac{5}{3}}, \label{high_s}
\end{eqnarray}
in the case of high $\beta$
and
\begin{eqnarray}
&&\tilde{\mathcal D}_{\rm m}(R)\approx C_{\rm m}(2/3)~[W_{\rm I}~(\hat{\mathcal P}_{\rm m}-\frac{1}{2}\epsilon~(\hat {\mathcal P}_{\rm m+2}+\hat{\mathcal P}_{\rm m-2}))
\nonumber\\
&&+W_{\rm L}{\sin^2\theta}\sum_{n=-\infty}^{\infty}[\hat{\mathcal P}_{\rm n}-\frac{1}{2}\epsilon~(\hat {\mathcal P}_{\rm n+2}+\hat{\mathcal P}_{\rm n-2})]~{A}_{\rm m-n}^{S}(0)]~R^{\frac{5}{3}},
\end{eqnarray}
in the case of low $\beta$. 
Here, the geometrical function for slow mode is
\begin{equation}
{A}_{\rm m-n}^{S}(\theta)=\frac{1}{2\pi}\int_0^{2\pi}d\psi e^{-i(m-n)\psi}\frac{\sin^2\psi}{1-\cos^2\psi\sin^2\theta}
\end{equation}
and other coefficients are the same as those of Alfv{\'e}n mode.

{\it Fast mode}: The multipole of synchrotron fluctuations is  
\begin{eqnarray}
&&\tilde{\mathcal D}_{\rm m}(R)\approx C_{\rm m}(1/2)~[W_{\rm I}~(\delta_{m0}-\epsilon\delta_{m2})+W_{\rm L}{\sin^2\theta}
\nonumber\\
&&\times({A}_{\rm m}^{F}(0)-\frac{1}{2}\epsilon~[{A}_{\rm m-2}^F(0)+{A}_{\rm m+2}^{F}(0)])]\hat{\mathcal F}_{0}R^{3/2}, \label{fast_high}
\end{eqnarray}
in the case of high $\beta$ and 
\begin{eqnarray}
&&\tilde{\mathcal D}_{\rm m}(R)\approx C_{\rm m}(1/2)~[W_{\rm I}~(\delta_{m0}-\epsilon\delta_{m2})+W_{\rm L}{\sin^2\theta}
\nonumber\\
&&\times({A}_{\rm m}^{F}(\theta)-\frac{1}{2}\epsilon~[{A}_{\rm m-2}^F(\theta)+{A}_{\rm m+2}^{F}(\theta)])]\hat{\mathcal F}_{0}R^{3/2},
\label{fast_low}
\end{eqnarray}
in the case of low $\beta$. 
Here, the geometrical function of fast mode, identical to slow mode, is expressed by
\begin{equation}
{A}_{\rm m-n}^{F}(\theta)=\frac{1}{2\pi}\int_0^{2\pi}d\psi e^{-i(m-n)\psi}\frac{\sin^2\psi}{1-\cos^2\psi\sin^2\theta}.
\end{equation}
In the following, we study the anisotropy using the ratio of the quadrupole moment ($m=2$) to the monopole moment ($m=0$).

\begin{figure*}
\centering
\includegraphics[width=1.0\textwidth,height=0.26\textheight]{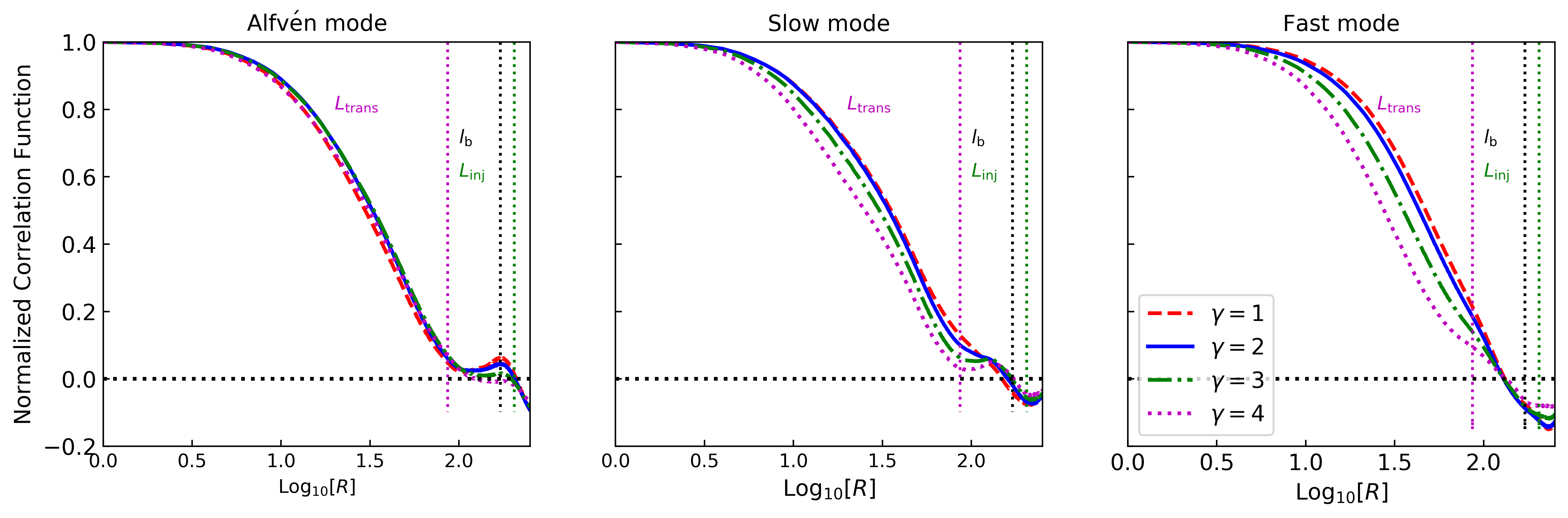}
\caption{The NCFs of synchrotron intensity for three modes decomposed by run5 listed in Table {\ref{table_1}}. For each panel, NCFs are plotted by changing magnetic field index $\gamma$. The green and magenta dotted vertical lines denote the injection and transition scales of turbulence, respectively. The black dotted vertical line denotes the correlation length of the magnetic field.
}\label{fig:ncf_three} 
\end{figure*}

\begin{figure*}
\centering
\includegraphics[width=0.9\textwidth,height=0.5\textheight]{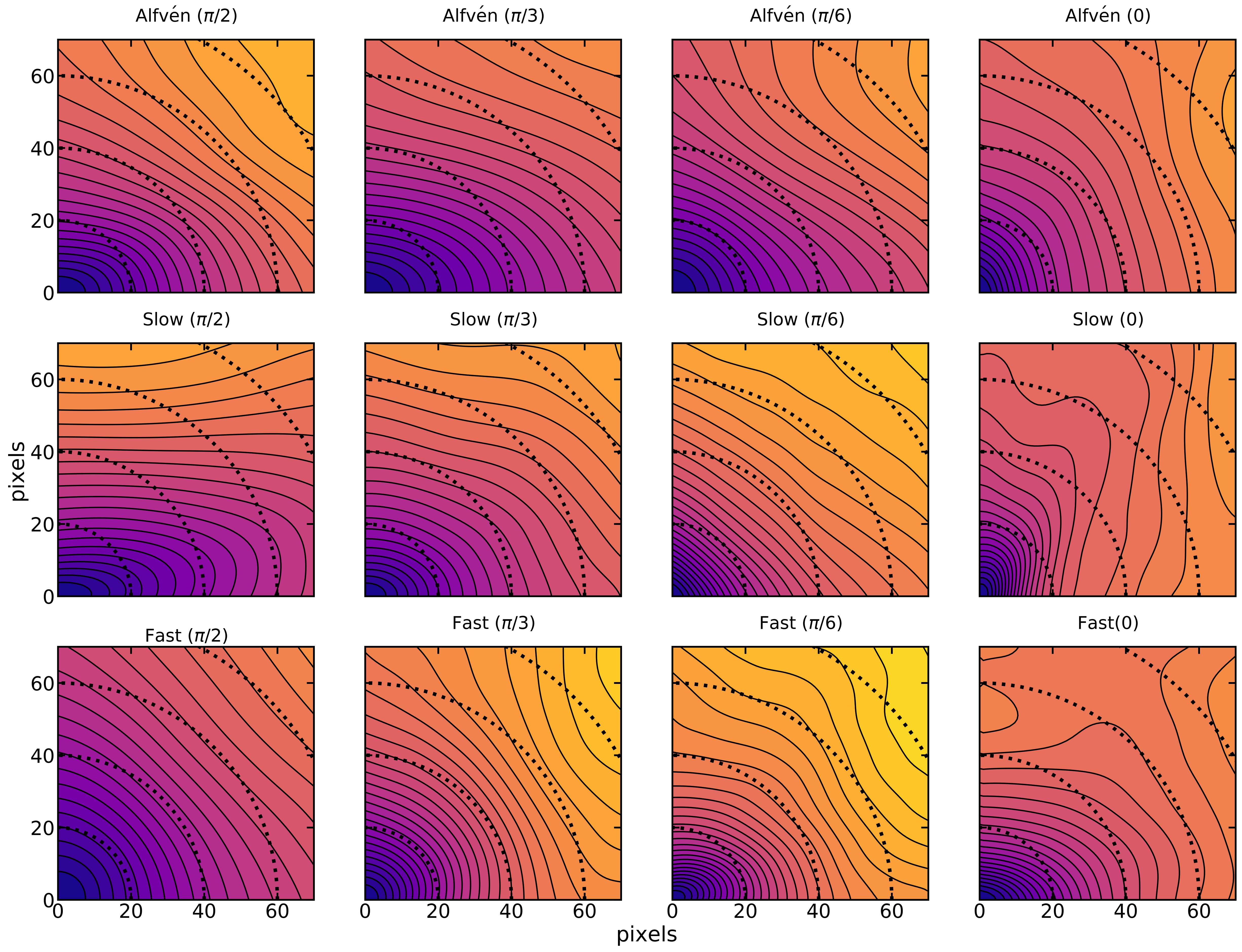}
\caption{Contour maps of normalized structure function of synchrotron radiation intensity for Alfv\'en, slow and fast modes, at the different angles between the mean magnetic field and the LOS. The dotted contour lines denote isotropy. The calculation is based on run5 listed in Table.} \ref{table_1}.
\label{fig:three_map} 
\end{figure*}

\subsection{Statistical Measures}

To numerically measure the anisotropy of MHD turbulence, we utilize the ratio of quadrupole moment to monopole one (QM):
\begin{equation}
{{\rm QM}({\bm R})}=\frac{\tilde{\mathcal D}_{2}}{\tilde{\mathcal D}_{0}}=
\frac{\int_0^{2\pi} e^{-2i\varphi} \tilde{\mathcal D}({\bm R},\varphi)~d{\varphi}}
{\int_0^{2\pi} \tilde{\mathcal D}({\bm R},\varphi)~d{\varphi}}, \label{eq:9}
\end{equation}
where ${\bm R}$ is a radial separation between two points, and $\varphi$ is the polar angle. The absolute value of $\rm QM$ characterizes the level of the anisotropy.\footnote{The QM plays a very important role in the theory of synchrotron gradients that is the basis of a way of magnetic field tracing (\citealt{Lazarian2017ApJ...842...30L}). } And the larger the absolute value, the more pronounced the anisotropy. The normalized structure function of radiation intensity ($Y=P$ or $I$) is expressed as 
\begin{equation}
\tilde{\mathcal D}=2(1-\tilde\xi), \label{eq:NSF}
\end{equation}
which is related to the normalized correlation function
\begin{equation}
\tilde\xi(\bm R)=\frac
{{\langle {\rm Y}({\bm X}){\rm Y}({\bm{X+R}})\rangle}-{\langle {\rm Y}({\bm X})\rangle}^2}
{\langle {\rm Y}({\bm X})^2 \rangle-{\langle {\rm Y}({\bm X})\rangle}^2}.\label{eq:NCF}
\end{equation}
where $\bm{X}=(x,y)$ is a two-dimensional position vector, and $\bm R$ denotes a separation vector between any two points on the plane of the sky (POS).

\section{MHD Turbulence Simulations and Analyses}
\subsection{Numerical scheme}
The evolution of the MHD turbulence is governed by the following set of equations
\begin{equation}
{\partial \rho }/{\partial t} + \nabla \cdot (\rho {\bm v})=0, \label{eq:den}
\end{equation}
\begin{equation}
\rho[\partial {\bm v} /{\partial t} + ({\bm v}\cdot \nabla) {\bm v}] +  \nabla p- {\bm J} \times {\bm B}/4\pi ={\bm f}, \label{eq:vel}
\end{equation}
\begin{equation}
{\partial {\bm B}}/{\partial t} -\nabla \times ({\bm v} \times{\bm B})=0,\label{eq:mag}
\end{equation}
with zero-divergence condition $\nabla \cdot {\bm B}$, and an isothermal equation of state $p=c_{\rm s}^2\rho$. Here, $t$ is the evolution time of fluids, ${\bm J}=\nabla \times {\bm B}$ the current density, and ${\bm f}$ a random driving force acting on large scale. 

Numerically, we use a third-order-accurate hybrid, essentially a non-oscillatory (ENO) scheme (CL02) to solve the above ideal isothermal MHD equations in a periodic box of size $2\pi$. More specifically, we combined two essentially non-oscillatory (ENO) finite difference schemes to mitigate spurious oscillations near shocks. When variables are sufficiently smooth, we use the third-order weighted ENO scheme (\citealt{Jiang1999J. Comput. Phys...150...561}) without characteristic mode decomposition. In the opposite case, we use the third-order convex ENO scheme (\citealt{Liu1998J. Comput. Phys...141...1}). The ENO schemes are generalizations of the total variation diminishing (TVD) schemes (\citealt{Harten1983J. Comput. Phys...49...357}). The latter typically degenerates to first-order accuracy at locations with smooth extrema while the former maintains high-order accuracy there even in multi-spatial dimensions. 

To maintain $\nabla \cdot {\bm B}=0$, we first solve for the potential $\varphi$ for the Poisson equation, $\nabla^{2}\varphi+\nabla \cdot {\bm B}=0$, with the updated magnetic field $\bm B$ obtained by the ENO scheme, and then we compute the corrected magnetic field as ${\bm B}_{\rm c} = {\bm B} + \nabla \varphi$, for which $\nabla \cdot {\bm B}_{\rm c}=0$. Furthermore, we use a three-stage Runge-Kutta method for time integration in units of the large eddy turnover time of $\sim L/ \delta V$. The magnetic field can be presented as $\bm{B}=\bm{B_{0}}+\bm\delta{B}$, i.e., a superposition of a regular magnetic field $\bm{B_{0}}$ ($\langle \bm B \rangle=\bm {B_{0}}$ due to $\langle {\bm \delta B}\rangle=0$) and a random/fluctuation magnetic field ${\bm \delta {B}}$.

\subsection{Generation of data cubes}
In our simulations, we set the mean magnetic field to be 1 or 0.1 along the $x$ axis. The turbulence is driven by a solenoidal driving force at the wavenumber $k=2.5$ (corresponding to the scale of $\sim 0.4L$) at Fourier space with a continuous energy injection and then it is transferred to the real volume space. When the simulation with the numerical resolution of $512^3$ reaches a statistically steady state at $t\sim 20$ in code units, we set the output to the primitive physical quantities we need, such as density\footnote{In this paper, we do not consider the density stratification and self-gravity effects,  which are not a part of the theoretical predictions that we test and not essential for most of the applications of the technique.} as well as individual components of the magnetic field and velocity. 

Based on the output 3D physical quantities and the parameters set at the initial moment, we obtain the typical parameters (e.g., the Alfv{\'e}nic Mach number $M_{\rm A}=\frac{V_{\rm L}}{V_{\rm A}}$ and sonic Mach number\footnote{Typically, the synchrotron emission regions correspond to the subsonic regime, but for the sake of completeness, we consider both subsonic and supersonic cases.} $M_{\rm s}=\frac{V_{\rm L}}{c_{\rm s}}$) listed in Table \ref{table_1} to characterize each run.
The former characterizes the strength of the magnetic field, and the latter reflects the compressibility, where $V_{\rm A}=|\bf B|/\sqrt{4\pi\rho}$ is Alfv{\'e}nic speed and $c_{\rm s}=\sqrt{P_{\rm g}/\rho}$ is the sound speed. The plasma parameter is described by $\beta=P_{\rm g}/P_{\rm m}=2M_{\rm A}^2/M_{\rm s}^2$, where $\beta>1$ represents gas-pressure-dominated turbulence and $\beta<1$ the magnetic-pressure-dominated one. 


\subsection{Analyses of data cubes}

The inertial range and strong turbulence range\footnote{When the assumption of the critical balance is satisfied, it corresponds to strong turbulence (see Section \ref{MHD_basic_theory} for more details).} can be shown by the correlation function, structure-function, and power spectrum. Since the calculation of the correlation and structure-function of the 3D magnetic field and velocity are time-consuming, we only use a power spectrum to determine these ranges. In Figure \ref{fig:power}, we present the power spectra of the magnetic field and velocity calculated at different turbulence regimes, where the dotted vertical lines represent injection, transition, and dissipation scales, respectively. The determination of these scales helps the calculation of the average QM. From this figure, we find that the power spectra of magnetic field and velocity satisfy $k^{-5/3}$ at four different turbulence regimes, but with different inertial ranges. 
Moreover, we find that the velocity spectra do not show obvious flattening in the vicinity of the dissipation range, so there is no bottleneck effect in the velocity spectra (\citealt{Falkovich1994}).

\subsection{Decomposition of MHD modes}
We input 3D data cubes of magnetic fields and decompose the data using the decomposition method. 
The decomposition of MHD turbulence is related to the following unit vector for Alfv{\'e}n, slow and fast modes (see CL03 for details)
\begin{equation}
\hat{\zeta}_{\rm f} \varpropto (1+\alpha + \sqrt{D})(k_{\perp} \hat {\bm k}_{\perp}) 
+(-1 + \alpha+ \sqrt{D})(k_{\parallel} \hat {\bm k}_{\parallel}), \label{eq:15}
\end{equation}
\begin{equation}
\hat{\zeta}_{\rm s} \varpropto (1+\alpha - \sqrt{D})(k_{\perp} \hat {\bm k}_{\perp}) 
+(-1 + \alpha- \sqrt{D})(k_{\parallel} \hat {\bm k}_{\parallel}), \label{eq:16}
\end{equation}
\begin{equation}
\hat{\zeta}_{\rm A} \varpropto -\hat{\bm k}_{\perp} \times \hat{\bm k}_{\parallel},\label{eq:17}
\end{equation}
where $D=(1+\alpha)^2-4 \alpha \cos^2\theta$ and $\cos\theta=\hat{k}_{\parallel} \cdot \hat{B}$. When projecting the magnetic field into $\hat{\zeta}_{\rm f}$, $\hat{\zeta}_{\rm s}$ and $\hat{\zeta}_{\rm A}$, we could obtain the magnetic field component for each mode.

\section{Results}

To generate synthetic observations, we calculate the synchrotron radiation intensity by the Equation (\ref{eq:I}), with the magnetic field from the simulation data and $\alpha=-1.0$. As for the calculation of synchrotron radiation, we ignore the fluctuation of electrons and only consider the fluctuation of the magnetic field.
For the synchrotron polarized radiation, its intensity is calculated by Equation (\ref{eq:PI}), with the assumption of the thermal electron density proportional to $\rho$ and we use $n_{\rm e}=\rho$ in the actual calculation of Faraday measure.
\subsection{Correlation of MHD turbulence}

The effect of the electron spectral index of cosmic rays on NCFs of synchrotron emissivity or intensity has been studied in \cite{Herron2016ApJ...822...13H} by numerical simulations. However, it was not right to adopt the expression applicable to isotropic turbulence (see Equation (7) of \cite{Herron2016ApJ...822...13H}) to study the correlation property of anisotropic turbulence. Here, we use the general formula, i.e., Equation (\ref{ncf}), to explore the dependence of NCFs of synchrotron emissivity on the spectral index in the case of anisotropic turbulence. At the same time, the dependence of spectral index is explored for different plasma modes in sub-Alfv{\'e}nic and subsonic turbulence regimes.

We first focus on the special case of NCFs of synchrotron emissivity, namely $\gamma=2$. It is explored in four turbulence regimes: sub-Alfv{\'e}nic and subsonic; sub-Alfv{\'e}nic and supersonic; super-Alfv{\'e}nic and subsonic; super-Alfv{\'e}nic and supersonic turbulence. The results obtained by Equation (\ref{ccc}) are shown in Figure \ref{fig:left_right}, where the upper and lower panels correspond to strong and weak magnetic field simulations, respectively. At the scale smaller than $L_{\rm trans}$, the turbulence becomes strong and the statistical relationship can be well represented.

This figure demonstrates that there is no expected consistency between the left and the right sides of Equation (\ref{ccc}), except for the presence of slight deviations in the case of sub-Alfv{\'e}nic and subsonic turbulence (see the upper left panel). Notice that Equation (\ref{ccc}) is derived under the assumption of only involving the fluctuation component perpendicular to the mean magnetic field, i.e., without the fluctuation along the mean magnetic field direction $B_{\rm x}$. Therefore, it is not surprising for the discrepancy to become more pronounced in the case of a weak magnetic field (lower panels) than the strong one (upper panels). In addition, there is a larger deviation for the high $M_{\rm s}$ regime, which may be due to the formation of shock. In the range from the injection scale $L_{\rm inj}$ (green line) to the transition scale $L_{\rm trans}$ (magenta line), the turbulence is weak (LV99; \citealt{Galtier2000JPlPh..63..447G}) and therefore the deviations from LP12 results are expected (see more details on the difference of turbulence in weak and strong regimes in \cite{Beresnyak2019tuma.book.....B}.

We then use the general expression, Equation (\ref{ncf}), to explore whether the varied spectral indices affect the statistics of the NCFs of synchrotron emissivity in the case of anisotropic turbulence. Figure \ref{fig:ncf_total} shows the NCFs of synchrotron emissivity as a function of radial separation at different spectral indices. We find that changes in the spectral index hardly affect the scaling index measured, except in the super-Alfv{\'e}nic and supersonic turbulence regime. Comparing the subsonic turbulence with the supersonic one, we find that the NCFs of synchrotron emissivity have a marginal dependence on the spectral index for the latter. It is not difficult to understand this phenomenon because the possible shock formation in the high $M_{\rm s}$ regime causes a deviation from the synchrotron statistics. 
In short, our study confirms that the correlation of synchrotron emissivity for an arbitrary index $\gamma$ of the magnetic field is linked to those for $\gamma =2$, namely,
\begin{equation}
{\rm CF}_{\gamma} (R)\approx \xi(\gamma) {\rm CF}_{\gamma=2} (R), 
\end{equation}
where $\xi(\gamma)=(\langle(B_{\perp}^{\gamma})^{2}\rangle-\langle B_{\perp}^{\gamma}\rangle^{2}/\langle B_{\perp}^{4}\rangle-\langle B_{\perp}^2\rangle^{2})$ is a factor that changes only with $\gamma$ and not with the separation $R$.
This is a practically important result in agreement with the theoretical prediction of LP12. 
However, it is important to stress that the above result does not hold in the super-Alfv{\'e}nic and supersonic turbulence regime. Our numerical confirmation of the theory opens ways for quantitative studies of synchrotron statistics for an arbitrary index of cosmic rays distribution.

Based on run5 listed in Table \ref{table_1}, we study the influence of electron indices on the NCF of synchrotron intensity arising from different MHD modes. The results of synchrotron statistics are shown in Figure \ref{fig:ncf_three}, where the range of spectral indices we consider corresponds to possible various astrophysical environments. 
As is shown in this figure, the electron spectral index has less effect on the results for the Alfv\'en mode, which is similar to the properties of pre-decomposition MHD turbulence and reflects that the Alfv\'en mode dominates the properties of MHD turbulence. However, the NCFs of synchrotron intensity for compressible slow and fast modes are completely dependent on the distribution of the electron index. They may be caused by the compressible nature of these two modes.

 \begin{figure*}
\centering
\includegraphics[width=1.0\textwidth,height=0.26\textheight]{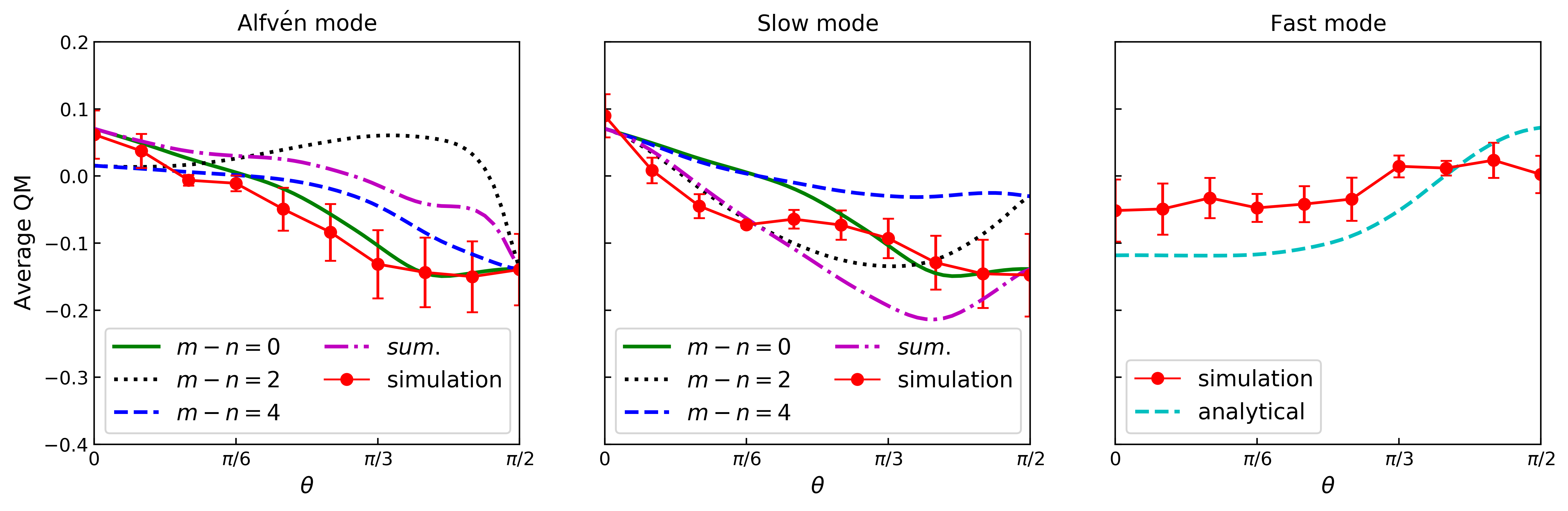}
\caption{{{
Average QM of synchrotron radiation intensity as a function of the angle $\theta$ between $B_{0}$ and the LOS for Alfv\'en, slow and fast modes.
The numerical results shown in the red lines are obtained by the run4 listed in Table \ref{table_1}.
The other lines represent the analytical results.
The legends $m-n=$0, 2, and 4 represent the order of the series expansion of the geometric function, and $sum.$ denotes a summation of the first three orders of the series expansion.
}}
}\label{fig:three_a} 
\end{figure*}

\subsection{Synchrotron Intensity Anisotropy Arising From Basic MHD Modes}

\subsubsection{Qualitative Analysis of Anisotropy}

To explore the effect of the angle between the mean magnetic field and the LOS on the anisotropy of MHD turbulence, we thus rotate the direction of the mean magnetic field in the $xoz$ plane along the $y$ axis. The angle $0^\circ$ denotes the LOS aligned with the $x$ axis and the angle $90^\circ$ for the LOS along the $z$ axis. Using statistics of synchrotron polarization intensity, \cite{Wang2020ApJ...890...70W} explored the anisotropy of compressible MHD turbulence. Notice that they only consider the special case on the mean magnetic field perpendicular to the LOS. Here, we focus on a more general case, that is,  we will investigate the dependence of the anisotropy for three modes on the angle between the mean magnetic field and the LOS based on synchrotron radiation intensity. 

\begin{figure*}
\centering
\includegraphics[width=1.0\textwidth,height=0.5\textheight]{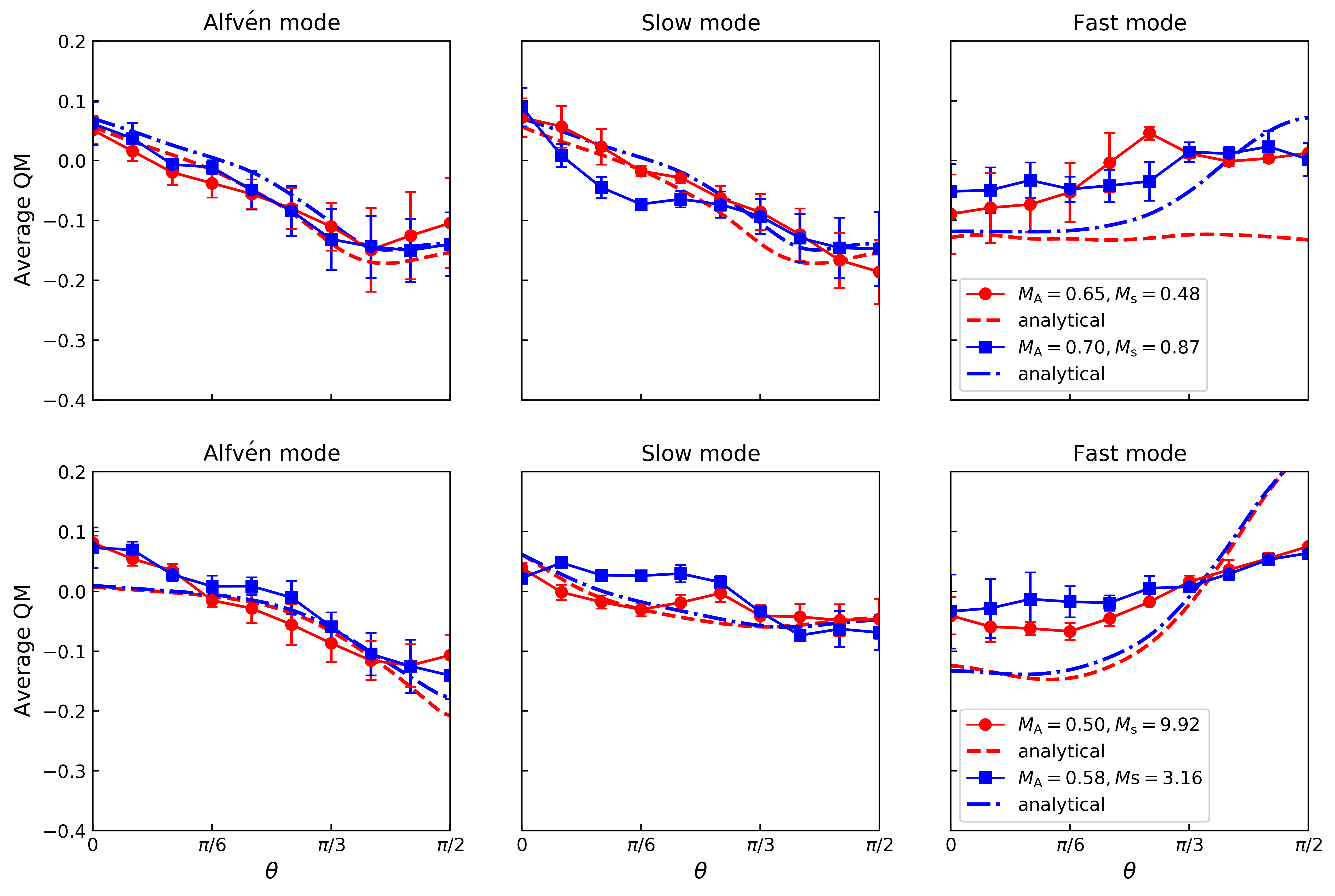}
\caption{Average QM of synchrotron intensity as a function of the angle $\theta$ between $B_{0}$ and the LOS for Alfv{\'e}n, slow and fast modes. The top and bottom rows represent the subsonic and supersonic simulations, respectively.
}\label{fig:three_s} 
\end{figure*}

We provide the contour maps of the normalized structure function of synchrotron radiation intensity for three modes in Figure \ref{fig:three_map}, which is based on the decomposed data of run5 listed in Table \ref{table_1}. From top to bottom, we display the contour maps for the Alfv{\'e}n, slow and fast modes, respectively. From left to right, the contour maps correspond to the angle $90^\circ$ ($\frac{\pi}{2}$), $60^\circ$ ($\frac{\pi}{3}$), $30^\circ$ ($\frac{\pi}{6}$), $0^\circ$, respectively. The angle $90^\circ$ represents the mean magnetic field perpendicular to the LOS and the angle $0^\circ$ is the opposite case.
As is shown in the upper and middle rows, the structures of contour maps are almost isotropic at a small angle for three modes. As the angle increases, the Alfv{\'e}n and slow modes exhibit significant anisotropic features while the fast mode remains almost isotropic. In the range of large angles, the structures of contour maps for three modes are similar to the earlier direct numerical simulation (CL02; CL03). It can be seen that synchrotron radiation statistics can efficiently reveal the information of the mean magnetic fields of the POS.

\subsubsection{Contribution of geometric function's order to $\rm QM$}\label{num_ana_com}

In the sub-Alfv\'enic and subsonic turbulence regimes, the numerical simulations of the average QM are compared with analytical predictions. Given that the anisotropy part of multipole involves different terms of the series expansion of the geometrical function for the Alfv{\'e}n and slow modes, we first explore which term of the geometrical function can match the anisotropy from simulation observations. For our purpose of studying $\rm QM$, the expansion order $m=0$ and 2 can be fixed for the monopole moment and quadrupole moment (see Equation (\ref{eq:9})). 
The contribution of the monopole moment ($m=0$) only comes from the zeroth order of the expansion of the geometrical function ($A_{0}^{A,~S}$ with $m-n=0$), while the contribution of the quadrupole moment ($m=2$) is related to  the zeroth, second and fourth orders of the series expansion of geometrical function ($A_{0, ~2,~4}^{A,~S}$ with $m-n=0,~2$ and 4).  Furthermore, once the expansion order of the geometric function is fixed, the expansion of power spectra will be determined via the unchanged $n$ (see Equations (\ref{amono}) and (\ref{high_s}) for details).

Figure \ref{fig:three_a} shows simulation results of the average QM as a function of the angle $\theta$ between the mean magnetic field and the LOS, in comparison with analytical results arising from different expansion orders of the geometrical function. 
For the simulation results, we calculate the average QM considering all the scales almost from 10 pixels to the transition scale (corresponding to the pixel listed in Table 1), namely in the strong turbulence range.
As for the analytical results, we consider the individual contributions from the zeroth, second, and fourth orders of the geometric function and summation of three orders, which corresponds to the legends $m-n=0,~2,~4$ and $sum.$ in Figure \ref{fig:three_a}, respectively. Note that there is no selection of parameter variations in the case of fast mode.

From left to right panels, we plot the average QM for the Alfv{\'e}n, slow and fast modes, respectively.
It can be seen that the average QM for Alfv{\'e}n and slow modes is increasing with the increase of the angle, while those of fast mode is almost unchanged with the increase of the angle. We find that for Alfv{\'e}n and slow mode the analytical results of the zeroth $m-n=0$ can match the simulations, and apart from some small deviations in the average $\rm QM$ amplitude, the overall trend of numerical and analytical results is also consistent for fast mode. The observed deviations as we add additional terms may mean that the expansion series in LP12 analytics requires using higher order terms, rather than limiting the expansion over the first two terms as it is done in the present study. 
In addition, we show the error bars corresponding to the variability with scale (taken from the standard deviation of the QM with the range of scales used for averaging). We find that the error bars for QM in the numerical simulation are consistent with the variation in the analytical calculation. Therefore, we believe that the error bars provide an adequate depiction of the reliability of the data.

Comparing with the analytical curves of QM provided in LP12, we find that at the angle $\theta=0$, the current analytical QM show non-zero values, which are caused by the stochastic deviations of the mean magnetic field from its original direction due to an external driving setting. Note that the analytical formulae of quadrupole moment include an important parameter, $\epsilon$, characterizing the level of turbulence anisotropy in the analytical formula (see Equation (\ref{aniscoeff})). In this regard, we have done a self-consistent treatment between the analytical theory and simulations, that is, the new non-zero $\epsilon$ value from the simulation is substituted into the analytical formulae, resulting in the non-zero QM values.

\begin{figure*}
\centering
\includegraphics[width=1.0\textwidth,height=0.26\textheight]{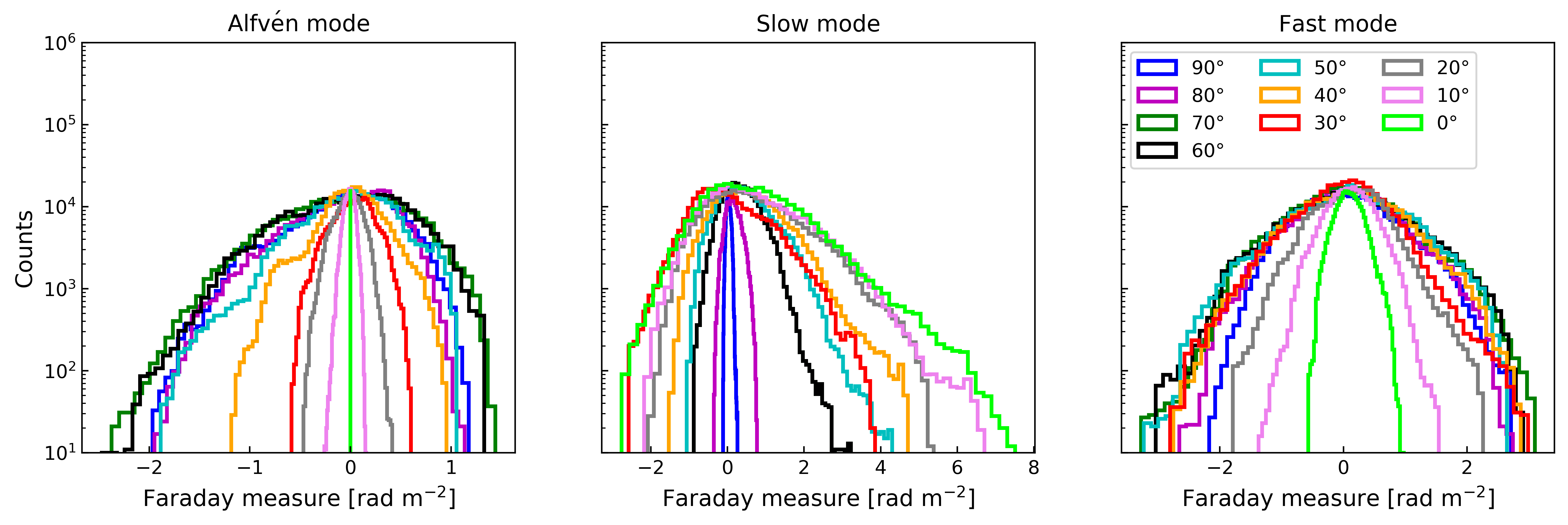}
\caption{Distributions of the Faraday measure for three modes at different angles between the mean magnetic field and the LOS.
} \label{fig:faraday} 
\end{figure*}

\subsubsection{Application to subsonic and supersonic turbulence regimes}\label{Appl}

In the section, we explore the relationship between the average QM and the angle for three modes in two different turbulence regimes, namely subsonic and supersonic turbulence regimes, shown in the top and bottom rows of Figure \ref{fig:three_s}, respectively. For the former, the analytical results are obtained by considering the contribution of the zeroth order of the series expansion of the geometrical function, while for the latter the contribution of the fourth order. The numerical results are calculated by the data cubes of Table \ref{table_1}.

In Figure \ref{fig:three_s}, it is shown that the anisotropy of Alfv{\'e}n and slow modes (left and middle columns) increases with the angle in different turbulence regimes, and the fast mode keeps almost isotropic for all the viewing angles due to approaching zero of the QM value. As is expected, the highest level of anisotropy appears in the case of the mean magnetic field perpendicular to the LOS. As is shown in the left and middle columns, the amplitude of the average QM for the Alfv{\'e}n mode is slightly larger than that of the slow mode at all angles. This reflects the fact that the Alfv{\'e}n mode dominates the properties of the slow mode while the fast mode with smaller amplitude is independent of these two modes. From top to bottom rows, it is shown that the absolute values of average QM for Alfv{\'e}n and slow modes in the subsonic turbulence regime are larger than those in the supersonic case, especially for slow mode. The reason may be that the formation of shock waves in a supersonic turbulence regime decreases the anisotropy of turbulence. Our simulations are in good agreement with the analytical prediction of LP12 for Alfv{\'e}n, slow modes in different cases. The deviations that we observe for fast modes may be due to their relatively small amplitude, which influences our decomposition. In particular, the effect of "leakage" of the modes, i.e. the contamination of one type of mode by other modes is considered in Yuen et al. (2022). 

Here, we would like to mention to interested readers that LP12 theoretical predictions did not consider the influence of the sonic Mach number $M_{\rm s}$ on the synchrotron fluctuation statistics. Our studies in this section promote the application of LP12 predictions in different turbulence environments. They show that LP12 theory can be applied to a variety of astrophysical conditions with different $M_{\rm s}$ for slow and Alfv{\'e}n modes, but its accuracy drops for fast modes with the increase of $M_s$. Fortunately, for practical applications this is not important, as most synchrotron-emitting media in astrophysical settings is the warm and hot media with low $M_s$.

\begin{figure*}
\centering
\includegraphics[width=1.0\textwidth,height=0.5\textheight]{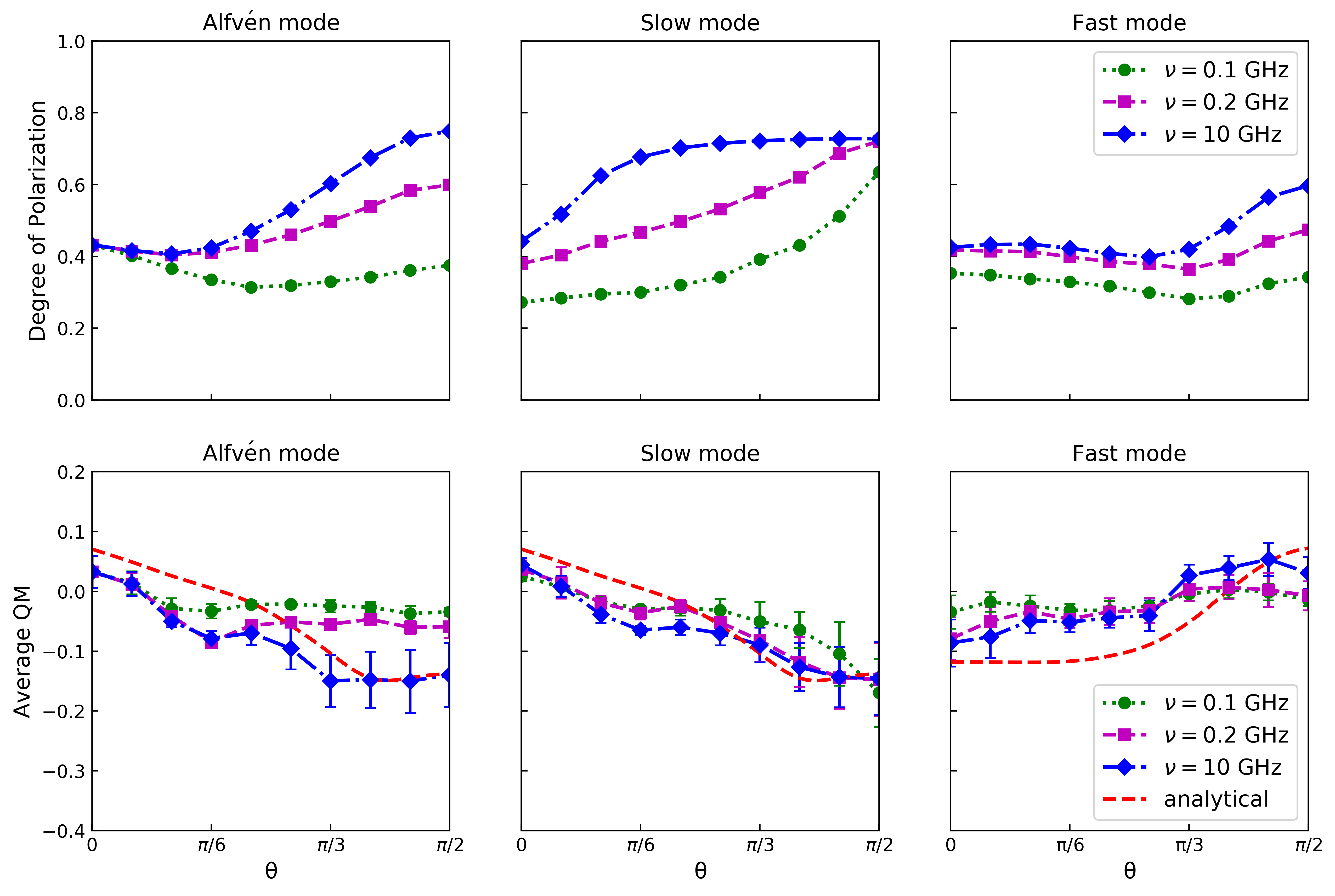}
\caption{The degree of polarization (upper panels) and average QM (lower panels) of synchrotron polarization intensity as a function of the angle $\theta$ for Alfv\'en, slow and fast modes at different frequencies shown in the top and bottom rows. The simulation results are based on run4 listed in Table \ref{table_1}, and the red dashed lines represent analytical predictions.
} \label{fig:three_sp} 
\end{figure*}

\subsection{Extention to Intensities of Polarized Emission}
Another important way to study MHD turbulence is through the analysis of synchrotron polarization information (see Section \ref{SecRad} for the calculation of the Stokes parameters). Note that the numerical simulation studies mentioned above are based on the theoretical basis of synchrotron radiation fluctuations of LP12. Here, we explore whether the relevant conclusions of the study of the synchrotron radiation fluctuation hold in the case of synchrotron polarized radiation. Moreover, we want to know to what extent the study of polarization statistics has an impact on the results in the presence of the Faraday rotation effect. In particular, we study whether measurements of the anisotropy are affected by the Faraday rotation effect in the different frequency regimes. 
We assume that the emission region is extended to be 1 kpc along the LOS, and thermal electron density and magnetic field strength are set as $n_{\rm e}\sim 0.01~{\rm cm}^{-3}$ (\citealt{Gaensler2008PASA...25..184G, Nota2010A&A...513A..65N, Lee2019ApJ...877..108L}) and $B_{\rm z}\sim 1.23~{\rm \mu G}$, respectively.

First, we show the distributions of the Faraday measure for three modes at different angles between the mean magnetic field and the LOS in Figure \ref{fig:faraday}, from which we clearly see that the Faraday measure for three modes follows non-Gaussian distribution (e.g., \citealt{Kierdorf2020A&A...642A.118K, Seta2021MNRAS.502.2220S}). As shown, the value of Faraday measure for three modes lies at different ranges: -2 to 1 $\rm {rad~m^{-2}}$ for Alfv{\'e}n mode, -2 to 8 $\rm {rad~m^{-2}}$ for slow mode, -3 to 3 $\rm {rad~m^{-2}}$ for fast mode. This reflects that the dispersion of the Faraday measure is different for the three modes and that the dispersion value of the slow mode is the largest. We find the dispersion of Faraday measure for Alfv{\'e}n and fast modes increases with increasing angle, while the opposite is true for that of the slow mode. In addition, the mean of the Faraday measure for three-mode at different angles is close to 0.

Figure \ref{fig:three_sp} shows the degree of polarization (top row) and the average QM (bottom row) arising from synchrotron polarization intensities as a function of the viewing angle, at different frequencies, i.e., $\nu=0.1, 0.2$ and $10~\rm GHz$. We see that the degree of polarization is increasing with both the increase of the angle and the frequency. This is because the increase of frequency and increased viewing angle that decreases the $B_{\parallel}$ component of the projected magnetic fields both reduce the Faraday rotation effect. At the same viewing angle and frequency, we find that fast mode has a lower degree of polarization than Alfv\'en and slow modes due to the isotropy of fast mode. The bottom row shows that the absolute values of the average QM of synchrotron polarization intensities for the Alfv{\'e}n and slow modes increase with the viewing angle $\theta$, while that of the fast mode presents small changes around the average QM equal to 0. The overall shape of average QM distributions for three modes has some changes at different frequencies, especially for Alfv{\'e}n and fast modes. 
These results reveal that in the wide frequency range considered, the Faraday rotation effect hardly hinders the measurement of the anisotropy for the case of high frequency.

\section{Discussion}

\subsection{Numerical tests of LP12 theory}

Our numerical study uses synthetic observations to test the predictions of LP12.
We focus on the effects of the relativistic electron spectral index and turbulence anisotropy on the statistics of synchrotron intensity and polarization. This study is essential to measure variations of synchrotron parameters, e.g., Faraday rotation (see \citealt{Haverkorn2008ApJ...680..362H, Waelkens2009MNRAS.398.1970W, Xu2016ApJ...824..113X}), Faraday tomography (\citealt{Burn1966MNRAS.133...67B, Haverkorn2017galactic}) as well as the new promising techniques of tracing magnetic field with synchrotron intensity (\citealt{Lazarian2017ApJ...842...30L}) and synchrotron polarization gradients (\citealt{Lazarian2018ApJ...865...59L, Carmo2020ApJ...905..130C}).

The prediction of LP12 regarding whether the correlation function of synchrotron emissivity depends on the spectral index in the case of the isotropic model has been explored by \cite{Herron2016ApJ...822...13H}, but they, unfortunately, used results obtained by the anisotropic turbulence to compare with the expression obtained for the toy model of isotropic turbulence.
In this paper, we test the general case that utilizes the anisotropic model to explore the dependence of NCF of synchrotron emissivity on the spectral index. Our results are not consistent with LP12 for both the case of super-Alfv{\'e}nic and supersonic turbulence.
The observed deviations are mainly caused by supersonic effects, e.g., shocks, that are not considered within the LP12 theoretical model.
Importantly, we explore the dependence of the normalized correlation function of synchrotron intensity of all three basic MHD modes on the spectral index and obtain expected results.

\cite{Wang2021MNRAS.505.6206W} demonstrated that the anisotropy of three modes can be better distinguished by QM vs. $R$ by only considering the mean magnetic field perpendicular to the LOS. In the paper, we explore the anisotropy of three modes from different angles. By average QM vs. $\theta$, we see that the overall trend of the three modes varies with angle, from which we can distinguish the three modes well. However, it is difficult to judge the three modes from the magnitude of the average QM because the overall variation for all the three modes is around $\lesssim 0.1$ in terms of average QM. This depends primarily on the mean magnetic field and is generally weakly dependent on the separation. For the former, the higher the magnetic field strength, the more significant the anisotropy, as shown in Figure 7 by Herron et al. (\citealt{Herron2016ApJ...822...13H}). For the latter, according to modern MHD turbulence theory, the smaller the scale is, the more significant the anisotropy. This has been confirmed in \cite{Wang2020ApJ...890...70W} by synchrotron radiation arising from compressible MHD turbulence.
In addition, when we generalize to synchrotron polarization intensity, frequency is also a factor affecting the amplitude of the average QM (see Figure \ref{fig:three_sp}).

\subsection{Ways of studying MHD turbulence}

The anisotropy of three modes has been explored by several studies. Initially, the anisotropy of three modes was obtained from the magnetic field, density, and velocity information of MHD turbulence (CL02; CL03). But this information is not directly available in real observations. Later, the study of anisotropy of three modes is extended to
synchrotron polarized radiation (\citealt{Wang2020ApJ...890...70W}) and velocity centroids (\citealt{Hernandez-Padilla2020ApJ...901...11H}). 
In addition to the anisotropy, the identification and relative contribution of different modes are important in different astrophysical environments. The former has been explored in Galactic turbulence (\citealt{Zhang2020NatAs...4.1001Z}).
The latter was studied for interstellar turbulence (\citealt{Hernandez-Padilla2020ApJ...901...11H}).

The structure-function map serves as an indicator for preferentially predicting the anisotropic structure. The anisotropy direction is related to the mean field over the POS and thus provides a way of tracing the magnetic field.  The structure-function map of Alfv{\'e}n mode in the case of angle $90^{\circ}$ shown in Figure \ref{fig:three_map} is the best proof, where the mean magnetic field is along the x-axis direction.
Compared with structure-function, the QM is more precise in both measurements of anisotropy and magnetic field tracing. 
Its magnitude reflects the level of anisotropy, and positive and negative values represent the direction of the anisotropic structures.
The two methods are synergistic for the study of properties of MHD turbulence, particularly in anisotropy and measurement of the magnetic field.

\subsection{Synchrotron statistics}

The radio synchrotron emission is primarily from the hot/warm, ionized diffuse medium, where the turbulence has a relatively low sonic Mach number $M_{\rm s}\le2$ (\citealt{Gaensler2011Natur.478..214G}). However, some environments, such as the regions of active galactic nuclei and supernova remnants interacting with the surrounding cold molecular cloud, could still have a large $M_{\rm s}$. This may be applicable to the colder, denser ISM, spectral lines related to velocity channel analysis and velocity correlation spectrum have been considered as the main probe (\citealt{Lazarian2000ApJ...537..720L, Lazarian2004ApJ...616..943L, Casanova2017ApJ...835...41G, Yuen2017ApJ...837L..24Y, Yang2021MNRAS.503..768Y}).

The development of synchrotron radiation techniques for measuring MHD turbulence is motivated by the massive amount of radio observational data that is currently or future available, such as LOFRA and SKA. The turbulence injection scale in spiral galaxies is expected to be of the order of the disk scale height, which is 100 pc or more (see \citealt{Chepurnov2010ApJ...714.1398C}). If observations resolve smaller scales, our technique is applicable. The resolution of new instruments constantly increases and thus we expect our technique to apply to more distant galaxies. At the same time, the limitations in terms of the required resolution are less strict for the Milky Way magnetic fields study. Thus the technique has a lot of present-day and future applications.

\section{Summary}

In this paper, the statistics of synchrotron intensity and synchrotron polarization fluctuation arising from the compressible MHD turbulence have been studied numerically with the results compared to the predictions of the analytical theory in LP12. Using the QM and correlation function, we explored the anisotropy of MHD turbulence and the influence of cosmic ray electron spectral index on synchrotron statistics, respectively.
The main results are briefly summarized as follows. 

1. Our simulations show the good correspondence of LP12 analytical expression of correlation statistics for a variety of magnetic field indices $\gamma$. In particular, we find that the approximation corresponding to $\gamma=2$ is adequate for describing the spatial variations of the fluctuations. 

2. The degree of the anisotropy for Alfv{\'e}n and slow modes is larger in subsonic simulations compared to supersonic simulations. The numerical simulations are in good agreement with the analytical results of LP12. 

3. As for the Alfv{\'e}n and slow modes, the degree of anisotropy increases with the angle between the mean magnetic field and the line of sight. The synchrotron statistics arising from the fast mode are almost isotropic. LP12 theoretical predictions regarding the anisotropy of plasma modes have been numerically confirmed. 

4. Analytical expressions related to synchrotron intensities have been generalized to the synchrotron polarization intensities. The results demonstrate that relevant expressions are still applicable, while the Faraday depolarization effect impedes the measurement of turbulence anisotropy.

\acknowledgments
We would like to thank the anonymous referee for the valuable comments that improved our manuscript. J.F.Z. thanks the support from the National Natural Science Foundation of China (grants No. 11973035 and 11703020) and the Hunan Province Innovation Platform and Talent Plan–HuXiang Youth Talent Project (No. 2020RC3045). R.Y.W. thanks the support from the Innovation Foundation of Xiangtan University (No. XDCX2022Y071). A.L. thanks the support
of NSF AST 1816234 and NASA TCAN 144AAG1967 and NASA
AAH7546 grants. F.Y.X. is supported by the Joint Research Funds in Astronomy U2031114 under a cooperative agreement between the National Natural Science Foundation of China and the Chinese Academy of Sciences.

\end{document}